# Harmonics-assisted optical phase amplification with a self-mixing thin-slice Nd:GdVO$_4$ laser operating in the self-induced skew cosh Gaussian mode

Kenju Otsuka[1] and Seiichi Sudo[2]

[1]TS$^3$L Research, 126-7, Tokorozawa, Saitama 359-1145 Janan
[2]Department of Physics, Tokyo City University, Tokyo 158-8557 Japan

*Abstract*— Harmonic-assisted phase amplification was achieved in a 300-µm-thick Nd:GdVO$_4$ laser in the self-mixing interference scheme. The key event is the self-induced skew cosh Gaussian (e.g., skew-chG) mode oscillation in a thin-slice solid-state laser with wide-aperture laser-diode pumping. The skew-chG mode was proved to be formed by the phase-locking of nearly frequency-degenerate TEM$_{00}$ and annular fields. The resultant modal-interference-induced gain modulation at the beat frequency between the two modal fields, which is far above the relaxation oscillation frequency, increased experimental self-mixing modulation bandwidth accordingly. Fifty-fold phase amplification was achieved in a strong optical feedback regime.

*Index Terms*— **Laser cavity resonators, Laser modes, Laser feedback, Laser velocimetry, Acoustooptic effects, Optical variables control, Optical variables measurement**

## I. Introduction

Following the demonstration of a flashlamp-pumped ruby laser by Maiman in 1960 [1], Kogelnik and Li published a monumental work on optical resonators in which they found that Hermite-Gaussian modes form in Fabry-Perot optical cavities as stable orthogonal transverse eigenmodes in laser resonators [2]. As for solid-state lasers, the lamp-pumped Nd:CaWO$_4$ laser and then the Nd:YAG laser were subsequently developed [3]. In 1974, the stoichiometric LiNdP$_4$O$_{12}$ (LNP) single crystal was developed by Yamada et. al. [4], while continuous wave laser-diode (LD) pumped microchip LNP lasers with short Fabry-Perot cavities smaller than 5 mm were demonstrated at room temperature in 1976 [5] and 1979 [6] at a threshold pump power of a few milliwatts. It was not until 1978, with the demonstration of a 1-W continuous wave laser-diode (LD) bar by Scifres et al. [7], that LD-pumped solid-state lasers began their rapid evolution that continues to this day [8].

In response to a variety of solid-state laser materials possessing high absorption coefficients at LD-pump wavelengths, e.g., LNP, Nd:GdVO$_4$ and Nd:YVO$_4$, we have, in the past few decades, studied LD-pumped microchip solid-state lasers with a focus on controlling transverse modes and nonlinear laser dynamics. Examples include selective excitations of transverse modes by using pump-beam manipulation [9, 10] and exploiting self-organized collective behavior in multimode lasers [11, 12]. Additionally, thinly sliced solid-state lasers with coated end mirrors (abbreviated as TS$^3$Ls) have been studied in a context related to the spatiotemporal dynamics of transverse modes in Fabry-Perot microcavities. The Fresnel number of a thin-platelet TS$^3$L cavity, NF = $a^2/l\lambda_l$ ($a$: aperture radius, $l$: optical cavity length, $\lambda_l$: lasing wavelength), is on the order of $10^2$–$10^3$ larger than those of conventional cavities. Large cavity Fresnel numbers enable lasing in a variety of transverse modes depending on the shape and spot size of the pump beam, i.e., by controlling the gain and thermally induced refractive index confinement of the lasing transverse modes as well as by controlling the actuated saturation type of optical nonlinearity inherent to TS$^3$Ls. These forms of lasing include vortex arrays originating from a higher-order Ince-Gauss mode and rectangular-type vortex arrays born from Hermite-Gauss modes in a 300-µm-thick LNP laser with shaped wide-aperture LD pumping [13] and Laguerre-Gauss modes arising from Ince-Gauss and Hermite-Gauss modes in a 1-mm-thick c-cut Nd:GdVO$_4$ laser with wide-aperture laser-diode pumping [14].

Starting with Otsuka in 1979 [15], TS$^3$Ls have been studied in a context related to highly sensitive self-mixing laser metrology. The self-mixing interference effect in TS$^3$Ls between the lasing field, $E_l$, and the weak optical field from the target, $E_s$, was found to be a simple self-aligned, cost-effective optical sensing technique that does not use sophisticated optical interferometers or highly sensitive electronics. Here, the laser acts as a high-efficiency mixer oscillator and shot-noise-limited quantum detector. The effective self-mixing modulation index is given by $m_e = 2\eta K$ ($\eta= |E_s/E_l|$: amplitude feedback ratio, $K = \tau/\tau_p$: fluorescence-to-photon lifetime ratio) and the power spectral intensity of the detected electrical self-mixing signal is proportional to $m_e^2$ [16-19]. TS$^3$Ls with large lifetime ratios have been used to make versatile self-mixing metrology systems with extreme sensitivity [20, 21].

On the other hand, dynamic changes in many physical quantities, including displacement, temperature, electrical and magnetic fields, can be transduced into changes in relative phase between light fields or wave functions. Therefore, most high-precision measurement tasks can be converted into measurements of a phase change in a specific physical process, and methods to amplify the phase are important for enhancing the resolution of phase measurements in metrology. Here, harmonics-assisted optical phase amplification has been demonstrated, wherein the relative phase difference between two polarization modes in a polarized interferometer is amplified coherently four times in cascaded second-harmonic generation processes [22]. Most recently, a new phase amplification method has been studied; it is based on the feedback-induced intracavity harmonic generation effect in a laser frequency-shifted feedback interferometer (the FIHG effect) employing a LD-pumped thin-slice Nd:YVO$_4$ laser [23]. It was reported that the relative phase change between the two arms of the interferometer is amplified by 11 times by the FIHG effect without assistance from any external harmonic generation. This figure exceeds the maximum amplification of around 10 that was experimentally obtained using a many-body entangled state in quantum optics [24,25].

On the basis of the above background, we examined ways of forming the lasing transverse mode inherent to TS$^3$Ls with wide-aperture LD pumping and their application to phase amplification by self-mixing modulation. In this paper, we describe skew-chG mode oscillations in a 300-μm-thick Nd:GdVO$_4$ laser with wide-aperture LD pumping for the first time, where an annular transverse field surrounding the central Gaussian field appears with increasing pump power by controlling the pump-beam diameter. A physical interpretation for the appearance of the annular lasing field is given in terms of the additional gain that appears around the preceding TEM$_{00}$ mode resulting from the transverse spatial hole burning effect of population inversions as well as the phase locking of TEM$_{00}$ and annular fields, which form skew-chG mode, associated with wide-aperture LD pumping. Harmonic-assisted phase amplifications in the skew-chG laser operation were demonstrated in the self-mixing laser Doppler velocimetry scheme as well as in the frequency-shifted optical feedback scheme using acousto-optic modulators (AOMs).

With increasing the optical feedback ratio, the strong amplitude modulation of coherent beat waves, which are generated through the modal interference of nearly frequency degenerate transverse modes, took place and extended the frequency bandwidth of harmonic-assisted phase amplifications far above the relaxation oscillation frequency. Fifty-fold phase amplification was achieved in the strong feedback regime.

## II. SELF-INDUCED SKEW-COSH GAUSSIAN MODE LASER OSCILLATION

### A. Wide-aperture LD Pumping and Slope Efficiency

The experimental setup with a Nd:GdVO$_4$ laser is shown in Fig. 1. A nearly collimated lasing beam from a laser diode (wavelength: 808 nm) was passed through an anamorphic prism pair to transform the elliptical beam into a circular one that was focused onto a 300-μm-thick laser crystal. One end surface was coated to be transmissive at the laser-diode pump wavelength of 808 nm (85% transmission) and highly reflective ($R_1 = 99.9\%$) at the lasing wavelength of 1063 nm. The other surface was coated to be $R_2 = 99\%$ at 1063 nm. Linearly polarized single longitudinal-mode emissions along the b-axis were observed in the entire pump power region, reflecting the fluorescence anisotropy. The pump-beam diameter was changed by shifting the laser crystal along the z-axis, as depicted in Fig. 1. The pump spot size, $w_p$, increased as the laser crystal was shifted away from the pump-beam focus along the z-axis (i.e., z > 0). The pump spot size varied from $w_p = 20$ μm (z = 0) to 85 μm (z = 2.5 mm). A pure TEM$_{00}$ mode oscillation was obtained at the threshold pump power, $P_{th} = 20$ mW, at $w_p = 20$ μm (z = 0), where the lasing spot size was $w_o = 30$ μm and the slope efficiency was $\eta_s = 24\%$, as shown on the left in Fig. 1. As $w_p$ increased, the threshold pump power gradually increased and the resultant lasing transverse mode exhibited a structural change in the near-field pattern. The slope efficiency increased to $\eta_s = 40\%$ regardless of the increase in threshold pump power to $P_{th} = 56$ mW for wide-aperture pumping at $w_p = 70$ μm because of the increase in the lasing mode volume as will be discussed in the following sections, *B* and *C*.

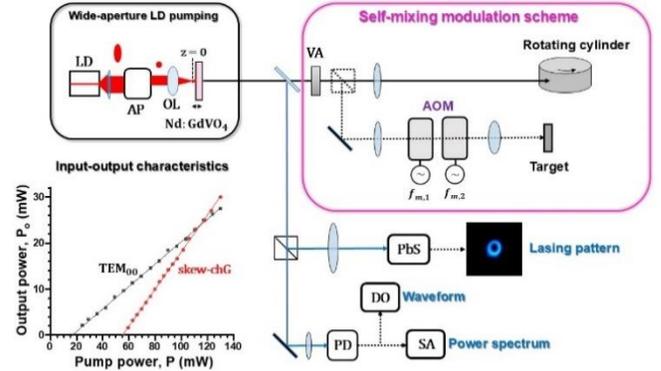

**Fig. 1.** Experimental apparatus of a thin-slice Nd:GdVO$_4$ laser with wide-aperture LD pumping, $w_p > w_o$, for harmonics-assisted phase amplification. Input-output characteristics for different pump spot sizes, $w_p \sim w_o$ and $w_p > w_o$. AP: anamorphic prism pair, OL: objective lens, VA: variable optical attenuator, PD: photo-diode, DO: digital oscilloscope, SA: spectrum analyzer.

The harmonics-assisted phase amplification was examined in two ways. One was self-mixing laser Doppler velocimetry by focusing the output beam onto a rotating Al cylinder by a 15-

cm focal-length lens placed 30 cm apart from the laser and the cylinder surface, where the laser was modulated at the Doppler-shift frequency, $f_D = 2v/\lambda_l$ (v: velocity along the laser axis). The other mixing modulation was performed using a pair of PbMoO$_4$ acoustic-optic modulators (AOMs; center frequency = 80 MHz) whose self-mixing modulation frequency is given by $f_M = 2(f_{m,1} - f_{m,2})$ depicted in Fig. 1.

### B. Lasing Beam Profiles

Near- and far-field lasing patterns were measured using a PbS phototube (Hamamatsu C1000) followed by a TV monitor and an intensity profiler. Typical results are shown in Fig. 2 with increasing the pump power for $w_p = 70$ μm. As can be seen in the movie, the pure TEM$_{00}$ mode appearing around the threshold pump power exhibited successive structural changes and the annular part increased with increasing pump power. It is noteworthy that the near-field intensity patterns in the present case are well fitted by the following skew cosh Gaussian (skew-chG) mode profile, where $I(r) = |E(r)|^2$:

$$E(r) = E(0)\cosh^n\left(\frac{rs}{b_w}\right)\exp\left[-\left(\frac{r}{b_w}\right)^2\right], \quad (1)$$

while the far-field patterns exhibited super-Gaussian-type intensity distributions of order $p$:

$$I(r) = I_0(0)\exp\left(-2\left|\frac{r}{w_0}\right|^p\right). \quad (2)$$

Here, $n$ is the order of skewness, and $s$ and $b_w$ are the skewness parameter and the beam width. Radial intensity profiles and fitting curves are shown on the lower row of Fig. 2, where the coefficient of determination is as high as $R^2 > 0.97$.

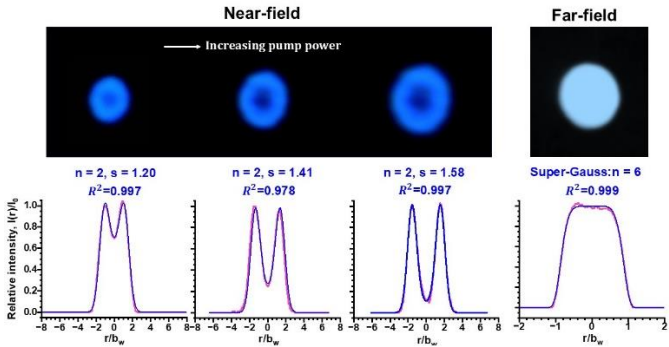

**Fig. 2.** Pump-dependent lasing near- and far-field patterns of thin-slice Nd:GdVO$_4$ laser with wide-aperture LD pumping. See Supplementary Material 1 (near-field) and 2 (far-field).

### C. Transverse Spatial Hole-Burning of Population Inversions

Let us give a plausible physical interpretation for the appearance of the annular gain associated with wide-aperture pumping. The spot sizes at the input and output mirrors, $w_1$ and $w_2$, and the effective focal length of the thermal lens, $f_T$, are given by [2, 26, 27]

$$w_i^2 = \left(\frac{\lambda n_0 l}{g_i}\right)\sqrt{\frac{g_1 g_2}{1 - g_1 g_2}}, \quad i = 1,2 \quad (3)$$

$$\frac{1}{f_T} = \left(\frac{lA}{2K_T}\right)\left[\left(\frac{dn}{dT}\right) + \alpha(n_0 - 1)\right]. \quad (4)$$

Here, $g_1 = 1 - \frac{n_0}{2f_T}$, $g_2 = 1$, $l$ is the effective thickness of the thermally induced lens, while $A$ is the heat generated per unit volume and time, $K_T$ is the thermal conductivity, $dn/dT$ is the thermal-optic coefficient of the refractive index, $\alpha$ is the coefficient of thermal expansion, and $n_0 = 1.972$ is the refractive index. Here, since $A$ increased as the pump power increased, the focal length was considered to decrease and the lasing beam spot size decreased accordingly. The pure TEM$_{00}$ lasing mode spot sizes at the crystal around the threshold pump power were measured to be 45~50 μm.

The TEM$_{00}$ spot sizes, $w_{1,2}$, were calculated using Eqs. (3) and (4) and the thermal constants of Nd:GdVO$_4$, $K_T = 11.7$ W/mK, $dn/dT = 4.7\times10^{-6}$/K, and $\alpha = 1.5\times10^{-6}$/K. They are plotted in Fig. 3(a), together with the average pump beam spot size, $w_p = 70$ μm. This figure indicates that wide-aperture pumping, i.e., $w_o < w_p$, is established for the case shown in Fig. 2. The near-field intensity profile observed at P = 120mW pump power is well fitted by a skew-chG function as shown in Fig. 3(b). A theoretical reconstruction of the skew ch-G lasing profile is presented in Fig. 3(c) by assuming the locking of two nearly degenerate TEM$_{00}$ and annular fields, namely $E_g$ and $E_a$, in the form of $E_g + irE_a$, where $r = (I_a/I_g)^{1/2}$ is the weighting number for field amplitude and $i$ implies the relative phase of $\pi/2$ [13].

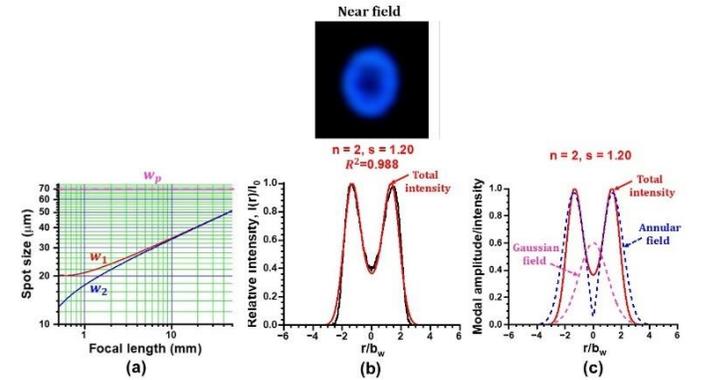

**Fig. 3.** (a) Calculated spot sizes at the crystal for the TEM$_{00}$ mode that arises from the thermal lens effect. (b) Example experimental near-field pattern and fitting using the skew ch-G intensity profile for pump power of P = 120mW. (c) Reconstruction of the skew ch-G profile assuming the Gaussian modal field and the annular field with the fixed relative phase of $\pi/2$.

Next, let us consider the transverse spatial hole-burning effect of population inversions. The pump rate of excited atoms and decreasing rate of excited atoms by lasing photons through stimulated emission are given by $\sigma_p I_p/h\nu_p$ and $\sigma_e I_{cir}/h\nu_o$, respectively. Here, $I_p$ and $I_{cir}$ denote the pump light and circulating lasing intensities, whereas $\sigma_e$ and $\sigma_p$ are respectively emission and absorption cross sections.

Assuming $w_p = 70$ μm, $w_o = 47$ μm, and spectroscopic data for Nd:GdVO$_4$: $\tau = 90$ μs, $\sigma_e = 7.6\times 10^{-19}$ cm$^2$ and $\sigma_p = 4.9\times 10^{-19}$ cm$^2$, calculated radial profiles related to the lasing intensity and population inversions are shown in Fig. 4:

(a) Pump-dependent intracavity circulating photon emission rate of the proceeding TEM$_{00}$ mode, which is given by $I_{cir} = I_s (I/I_{th} -1) \cong P_o/(\pi w_o^2)\ln(-R_2)$, where $I_s = h\nu_o/\sigma_e\tau$ is the emission saturation intensity, $I_{th}$ is the threshold pump intensity and $P_o$ is TEM$_{00}$ pump-dependent output power component as estimated from the input-output characteristics in Fig. 1 and modal field reconstructions as shown in Fig. 3(c) [28].

(b) Remaining atom excitation rate (i.e., remaining population inversions) in the presence of the preceding TEM$_{00}$ mode for various pump intensities.

(c) Effective atom excitation rate for the annular region defined as the hole depth, $10\log(N_r/N_{TEM})$, where $N_r$ is the remaining atom excitation rate in Fig. 4(b). The annular gain region spreads outward with increasing pump power and reproduce experimental results in Fig. 2.

As the pump power increases, depletion of the population inversion (i.e., transverse spatial hole burning) takes place. Here, the spatial integral of the remaining population inversion, $V = \iiint N(x,y,z)\,dv$, was found to coincide with the threshold value for $I_{cir} = 0$ for all pump powers within 3% error. This strongly implies that the population inversion density is kept at the threshold value under the lasing condition for the preceding TEM$_{00}$ mode obeying laser theory. While, the remaining population inversion shown in Fig. 4(b) is expected to give an additional effective gain for annular mode to coexist with TEM$_{00}$ mode with increasing the pump power so that almost all the population inversion contributes to lasing by wide-aperture pumping. Therefore, the lasing mode volume increases and the higher slope efficiency is expected as compared with the TEM$_{00}$ mode. These double-peaked population inversion distributions, which are fitted by a sum of decentered Gaussian distributions with $R^2>0.99$, are considered to make use of most of the population inversion to produce annular Gaussian-type emissions via the gain guiding effect [29].

In fact, the additional gain for an annular lasing field determined from the amplitude ratio of two fields at $P = 120$mW shown in Fig. 3(c) is 2.07dB and it coincides well with 1.94 dB evaluated from Fig. 4(c).

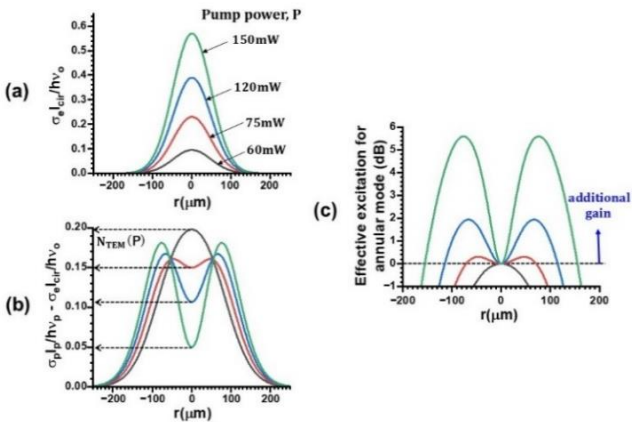

**Fig. 4.** (a) Relative pump-dependent circulating photon emission rate of TEM$_{00}$ mode. (b) Relative atom excitation rate in the presence of TEM$_{00}$ mode. (c) Effective atom excitation for annular fields.

On the other hand, the population-inversion-dependent refractive index variation, $\Delta n$, can be expressed as [30, 31]

$$\Delta n = \left(\frac{2\pi}{n_0}\right) f_L^2 N_e \Delta\alpha, \qquad (5)$$

Here, $f_L = (n_0^2 + 2)/3$ is the Lorentz local-field correction factor, $N_e$ denotes the excited-state ion population and $\Delta\alpha = a_e - a_g$ is the difference in polarizability of active ions in the metastable and ground states. Usually, $N_e$ is, to first order, proportional to the pump intensity, $N_e \approx N_T(I_p/I_{s,a})$, where $N_T$ is the terminal-state ion population, $I_{s,a} = hc/\lambda_p\sigma_p\tau$ is the absorption saturation intensity at the excitation wavelength, $\lambda_p$. Since $N_e$ is proportional to $I_P$, Eq. (5) can be written in terms of the pump-intensity-dependent refractive index change, $\Delta n = n_2' I_P$: [30, 31]

$$n_2' = \left(\frac{2\pi}{n_0}\right) f_L^2 N_T \Delta\alpha / I_{s,a}, \qquad (6)$$

For Nd:GdVO$_4$ crystals, the value of $\Delta\alpha$ is unknown, but the annular emission is expected to join with the pure Gaussian emission (which obeys Eqs. (3)-(4)) because of the pronounced modal gain $G(I_p)$ resulting from gain guiding and refractive index confinement with wide-aperture pumping, i.e. $w_p > w_o$, in accordance with Fig. 4(c) and Eq. (6).

### III. SELF-MIXING MODULATIONS IN SKEW-COSH GAUSSIAN MODE LASER

#### A. Phase Amplifications in Weak Feedback Regime

In this section, we address the key issue in the nonlinear dynamics of such skew-chG mode lasers subjected to self-mixing interference modulations by employing two methods, self-mixing laser Doppler velocimetry and frequency-shifted feedback using acousto-optic modulators, as shown in Fig. 1. Here, self-mixing interference metrology works through the intensity modulation effect of a laser due to interference between the lasing and feedback fields. In short, the self-mixing laser acts both as a mixer-oscillator and highly sensitive detector of the signal from the target. The resultant optical sensitivity has been shown to be enhanced in proportion to the square of the fluorescence-to-photon lifetime ratio, $K = \tau / \tau_p$, which reaches an order of $10^5$–$10^6$ in state-of-the art TS$^3$Ls. The photon lifetime in the present Nd:GdVO$_4$ laser was determined by measuring the dependence of the relaxation oscillation on the excess pump rate: $f_{RO} = (1/2\pi)[\sqrt{(w-1)/\tau\tau_p}$, $w = P/P_{th}$, as shown in Fig. 5(a). Assuming $\tau = 90$ μs, $\tau_p$ of 20 ps was attained, yielding $K = 4.5 \times 10^6$.

Self-mixing interference effects in a skew ch-G laser subjected to subharmonic modulations, $f_M = f_{RO}/2$, have been found to depend critically on the feedback ratio from the target. In the case of the self-mixing laser Doppler velocimetry scheme in Fig. 1, the effective intensity feedback ratio from the same rotating Al cylinder for the pure TEM$_{00}$ operation was estimated to be $\eta = -83$ dB from the correspondence between the experimental and numerically reproduced power spectral intensities of the Doppler signal at f$_D$ [16].

By inserting a variable attenuator with a roundtrip attenuation of T$_A$ $\cong -9$ dB in Fig. 1, the deep period-2 modulation took place as shown in Fig. 5(b). The similar periodic-2 modulation was achieved in the AOM scheme as shown in Fig. 5(b) by controlling the feedback condition from the target (i.e., Al plate) in Fig. 1 similarly to the LDV scheme mentioned above.

of the system, C$_N$ represents the intensity output coefficient of the Nth harmonic, which is related to the external optical loss. [23]. They demodulated $\Delta\varphi$ through the lock-in amplifier and proved that the dependence on N$\Delta\varphi$ instead of $\Delta\varphi$ allows us to achieve phase super-resolution measurement of $\Delta\varphi$, because the phase oscillation is N times faster than the original phase change and the unwrapped phase change $\Delta\varphi_u$ is linearly proportional to the optical path change $\Delta L$ as $\Delta\varphi_u = N(2\pi/\lambda)\Delta L$ [23].

### B. Intermediate Feedback Regime

The additional gain shown in Fig. 4(c) as well as pump-intensity-dependent refractive index given by Eq. (6), which are determined by the thermal lens effect, i.e., Eqs. (3)-(4), excite the annular mode around preceding TEM$_{00}$ mode. The refractive index (equivalently, laser cavity length) for the annular mode increases with increasing the pump power and the lasing frequency is expected to be shifted slightly from that of the central TEM$_{00}$ mode accordingly. With increasing a feedback ratio, the periodic oscillation waveforms shown in Figs. 5 and 6 tend to be modified, where self-mixing modulation originating from the two fields in Fig. 3(c) in the skew-chG profile are superimposed.

An example output waveform obtained by the self-mixing LDV scheme and the corresponding power spectrum are shown in Fig. 7(a), at T$_A \cong -6$ dB. The power spectrum has a secondary peak around 24 MHz whose envelope has a Gaussian profile. The enlarged peculiar total output waveform is shown at the top of Fig. 7(b). Filtered output waveforms below and above 17 MHz are shown in the middle of Fig. 7(b), where the total output in the lower-frequency band, I$_L$(t), exhibits a period-2 like periodic waveform reflecting the self-mixing modulation due to the coexisting modal components similarly to Fig. 5, whereas the total output in the higher-frequency band exhibits periodic oscillations at $\Delta\nu_B = 24$ MHz, whose envelope is modulated by the periodic waveform in the lower-frequency band.

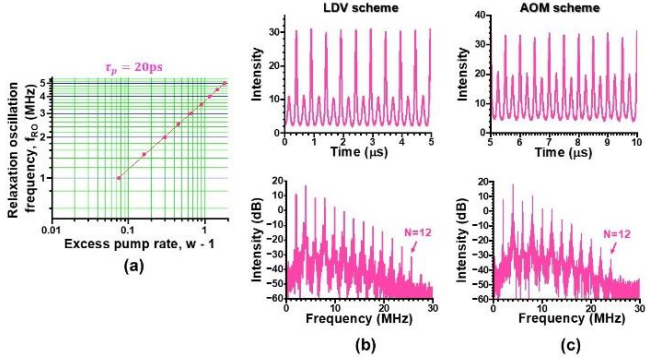

**Fig. 5.** (a) Dependence of f$_{RO}$ on the excess pump rate, w − 1. (b), (c) Waveforms and corresponding power spectra under the weak feedback regime. Pump power, P = 120 mW.

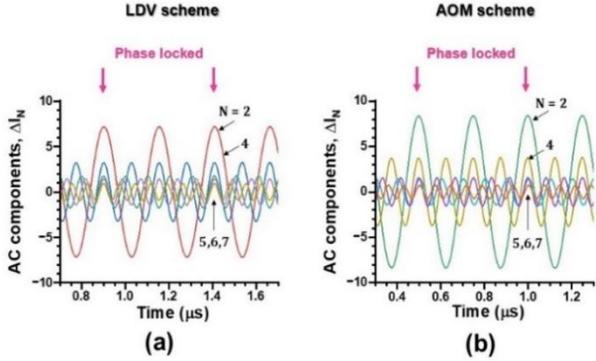

**Fig. 6.** Example harmonic waves for LDV and AOM self-mixing schemes.

Period-2 pulsations inherent to phase amplification are considered to appear through coherent superposition of N harmonic waves. Figure 6 shows example harmonic waves, which were filtered from the time series shown in Figs. 5(b) and 5(c) with a bandwidth of 100 kHz. The phases of the Nth-harmonic waves coincide with that of the fundamental wave every N periods. There are excellent phase correlations up to N = 12 in both plots. The observed phase-locking is direct experimental evidence supporting Tian and Tan's assertion that the intensity modulation (i.e., AC component) of the N-th harmonic from the self-mixing laser, $\Delta I_N$, is given by

$$\Delta I_N/I \propto C_N \cos[N(2\pi f_D t - \varphi_0) + N\Delta\varphi]. \quad (7)$$

Here, $\Delta\varphi$ is the relative phase change between the two arms in the self-mixing laser interferometer, $\varphi_0$ is the initial fixed phase

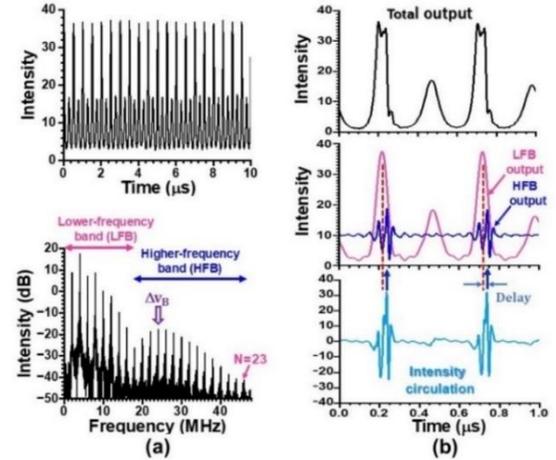

**Fig. 7.** Response of skew ch-G laser subjected to self-mixing LDV. The total output in the lower- and higher-frequency band is shown in the middle plot in (b). P = 120 mW.

In the present experiment, the gain (stimulated emission) modulation is considered to be brought about at a beat frequency, $\Delta\nu_B$, through modal interference of phase-locked transverse modal fields shown in Fig. 3(c) in the form of $BN_0(\iint \tilde{E}_g \tilde{E}_a^* dxdy + c.c)$, where $\tilde{E}_g$ and $\tilde{E}_a$ are the preceding $TEM_{00}$ field and the self-excited disturbing annular field whose all components seem a superposition of $LG_{0,\pm 1}$ modes of opposite azimuthal order which does not carry optical angular momentum [32]. $B$ is the stimulated emission coefficient and $N_0$ is the population inversion density. Instead of the complete transverse mode locking, the present skew ch-G lasing pattern is formed of nearly frequency-degenerated $TEM_{00}$ and annular fields with the fixed relative phase of $\pi/2$ as discussed in II-C, where the laser is modulated by the beat frequency at $\Delta\nu_B$ and might suffer the enhanced self-mixing modulation at $f_M \gg \Delta\nu_B$ as well with increasing the feedback ratio. As a result, the power spectrum shown in Fig. 7(a), arises around $\Delta\nu_B$ and forms the higher-frequency band, where the subharmonic resonance of $f_M/\Delta\nu_B = 12$, is established among the coexisting modal fields.

To clarify the interplay between these frequency bands, we carried out statistical analyses on total output waveforms belonging to the lower- and higher-frequency bands, i.e., $I_L(t)$ and $I_H(t)$, based on the observable quantity of intensity circulation given by [33, 34]

$$I_{L,H} = I_{L\to H} - I_{H\to L} = I_L(t)\dot{I}_H(t) - \dot{I}_L(t)I_H(t). \quad (8)$$

The calculated intensity circulation is shown in the bottom plot of Fig. 7(b). Here, it can be seen that $I_H(t)$ and $I_{L,H}(t)$ are correlated and $I_H(t)$ exhibits peaks in accordance with the intensity transfer from the lower to higher-frequency band, i.e., $I_{L,H}(t) > 0$, as depicted by the arrows. This implies that the intensity transfer occurs from the lower to the higher band just after the stronger peak indicated by the dashed line appears in $I_L(t)$ and the total intensity $I_H(t)$ reaches the maximum value within the periodic bursts at $\Delta\nu_B$.

Next, we examined the phase relationships among coexisting periodic waveforms of different N within the lower- and higher-frequency bands. The phases of the N-th harmonic wave in the higher-frequency band shown in Fig. 8(a) locked according Eq. (7). The total output belonging to the higher-frequency band exhibited antiphase dynamics against the harmonics in the lower-frequency band, where the bottom intensity of the harmonics in the higher-frequency band as shown in Figs. 8(b) and 8(c).

After quite a short time delay depicted in Fig. 7(b) and Fig. 8(c), harmonic waves in the higher-frequency band are phase-locked as indicated by the red arrows in Fig. 8(b). In other words, the gain transfer from the lower to higher frequency band corresponding to the intensity transfer, depicted by the blue arrows in Figs. 7(b) and Fig. 8(c), triggers phase locking among all the harmonics in the higher-frequency band after a slight time delay on the order of $1/(2\Delta\nu_B)$. Consequently, the harmonic waves in both frequency bands obey Eq. (7), indicating the subharmonic resonance of $f_M = \Delta\nu_B/12$ in this case via a slight time delay.

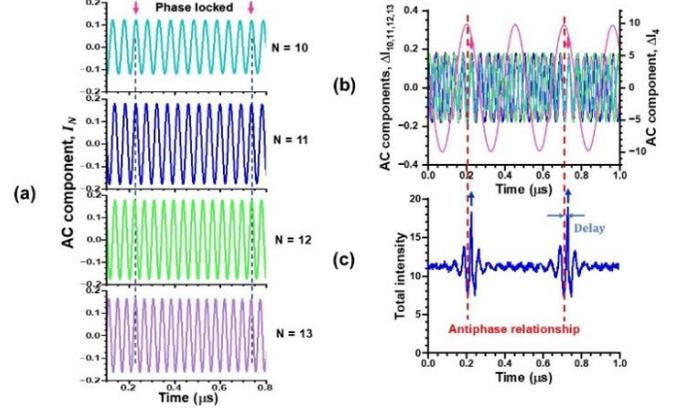

**Fig. 8.** (a) Phase-locking among harmonics in the higher-frequency band. (b) Phase relationship between harmonics in different frequency bands. (c) Total intensity waveform belonging to the higher-frequency band. P = 120mW.

Similar nonlinear dynamics leading to the peculiar phase locking behaviors illustrated in the self-mixing laser Doppler velocimetry shown in Fig. 8 were produced by the frequency-shifted AOM feedback scheme as shown in Fig. 9.

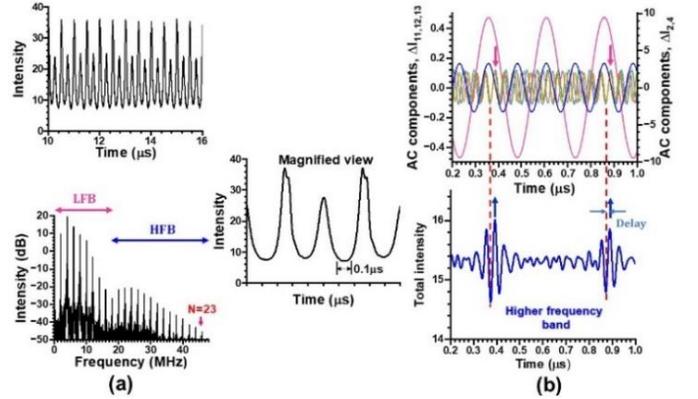

**Fig. 9.** Subharmonic resonance observed in the AOM feedback scheme. (a) Oscillation waveform together with a magnified view and the corresponding power spectrum. (b) Typical waveforms of the N-th harmonics and their phase relations with the total intensity waveform belonging to the higher frequency band. P = 120 mW.

*C. Strong Feedback Regime*

When the feedback was increased further, $T_A \cong -3dB$ self-mixing signals exhibited more complicated behavior, featuring low-frequency envelope modulations. An example result obtained in the AOM feedback scheme is shown in Fig. 10.

Here, the second peak is shifted downward to 20 MHz and its harmonic peak appears at 40 MHz in the power spectrum. Despite such a nonlinear effect with increased feedback, harmonics-assisted phase amplifications up to N = 30 were achieved, as shown in Fig. 10(c).

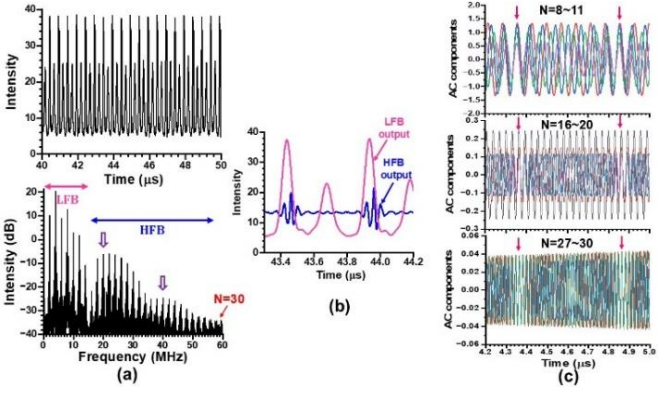

**Fig. 10.** (a) Self-mixing modulation waveform and the corresponding power spectrum in strong feedback regime. (b) Magnified view of outputs in low- and high-frequency bands. (c) Phase-locking among harmonics up to N = 30. P = 120 mW.

Finally, let us show a fifty-fold phase amplification observed by removing VA in LDV scheme. Results are shown in Fig. 11, where the pump power was decreased to P = 75 mW and the modulation frequency was set as $f_M = f_{RO}/2 = 1$ MHz. In this case, the lasing profile approached TEM$_{00}$ as shown in Fig. 2 and the second peak $\Delta\nu_B$ was shifted downward reflecting a decreased pump-intensity-dependent refractive index change given by Eq. (6), where harmonics up to $3\Delta\nu_B = 38$ MHz appeared. The phase amplification up to N = 50 and essentially the same nonlinear dynamics as Figs. 7-10 were observed.

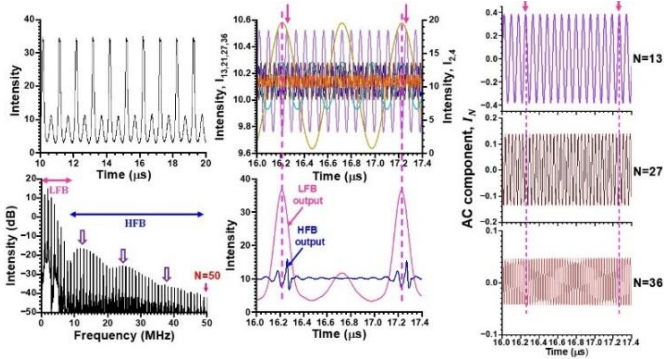

**Fig. 11.** Fifty-fold phase amplification with the decreased pump power in the absence of VA. $f_M = f_{RO}/2 = 1$ MHz. P = 75 mW.

## V. STATISTICAL PROPERTIES OF COLLECTIVE DYNAMICS IN LOWER- AND HIGHER- FREQUENCY BANDS

The small-signal modulation bandwidth beyond the relaxation oscillation frequency was studied in the context of a monolithic twin-ridge laterally coupled diode laser, where single- and double-lobed lasing modes coexist with split lasing frequencies [35]. The study experimentally showed that these modes exhibit mode locking and a lateral coupling resonance frequency arises above the relaxation oscillation frequency, similarly to our skew-chG mode operation involving two transverse lasing fields.

The significant dependence of the self-mixing subharmonic modulation effect on the feedback coefficient in the skew ch-G mode laser described in Section III can be interpreted in terms of the lateral coupling of Gaussian and annular fields within the laser. An example intensity power spectrum in the free-running condition, which follows the predictions of small signal analysis over a wide frequency range beyond the relaxation oscillation frequency [36], is shown in Fig. 12(a). The power spectrum is fitted by a Lorentzian profile in the lower frequency band and a Gaussian profile in the higher frequency band. This spectral nature is inherited to the harmonics-assisted phase amplification in the intermediate feedback regime, as shown in Figs. 12(b) and 12(c), where the power spectral intensity of harmonics in the lower frequency band obeys a Lorentzian distribution while that in the higher frequency band follows a Gaussian distribution.

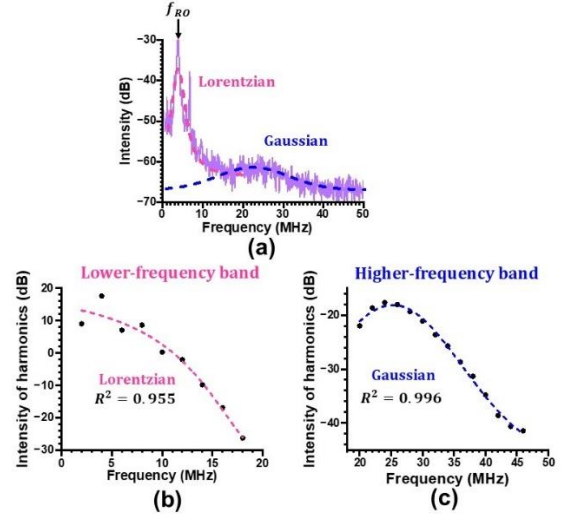

**Fig. 12.** (a) Power spectrum in skew ch-G operation under the free-running condition. (b), (c) Power spectral intensity of harmonics. P = 120mW.

The peak forming the higher-frequency band might be directly related to the lateral-coupling resonance frequency $\Delta\nu_B$ around a beat frequency between the TEM$_{00}$ and annular modal fields. Indeed, the broad Gaussian peak around $\Delta\nu_B$ decreased in intensity when the pump power was decreased such that the lasing pattern approached the pure TEM$_{00}$ mode oscillation.

Finally, it is interesting that there is a non-trivial correlation between the histogram and power spectra in the present laser system. Figure 13 shows the histograms (i.e., intensity probability distribution) of the output waveforms in the lower and higher frequency bands. They are fitted quite well by Lorentzian and Gaussian distributions, respectively, and exhibit a strong correlation with the power spectra shown in Figs. 12(b) and 12(c). The power spectral nature shown in Fig. 12(a) in the present skew ch-G mode operation with lateral-coupling resonance of coexisting modes seems to play a crucial role in organizing collective dynamics.

The same histogram nature was also established for Figs. 10-11 in the strong feedback regime, i.e., Lorentzian for the lower-frequency band output and Gaussian for the higher-frequency band output.

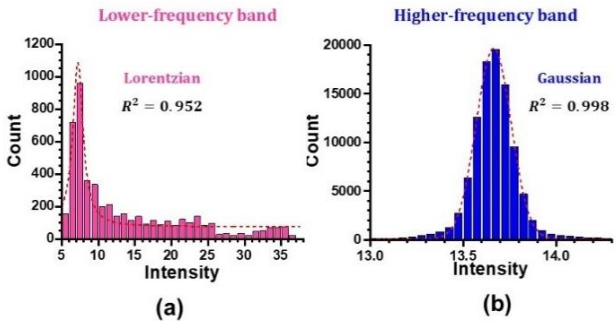

**Fig. 13.** Histogram for output intensities for (a) lower-frequency band and (b) higher-frequency band.

. SUMMARY AND OUTLOOK

Self-induced skew cosh Gaussian mode oscillations were observed in a 300-μm-thick Nd:GdVO$_4$ laser with coated end mirrors during wide-aperture laser-diode pumping. This is the first report on high slope efficiency skew-chG mode laser oscillations with a large fluorescence-to-photon ratio, which ensures the highly-sensitive self-mixing laser metrology.

The observed skew ch-G mode was proved to be formed by the phase locking of nearly frequency-degenerate TEM$_{00}$ and annular modal fields with the fixed relative phase of π/2. It has been clarified that the spatial hole burning effect of population inversions by the preceding TEM$_{00}$ mode provides the annular mode with an effective gain and the modal-interference-induced modulation takes place at the beat frequency, which is far above the relaxation oscillation frequency.

We carried out harmonics-assisted phase amplification experiments with the present skew ch-G mode laser with self-mixing subharmonic modulation both in laser-Doppler velocimetry and frequency-shifted AOM feedback schemes and examined the dependence of the nonlinear dynamics on the feedback ratio. With increasing the feedback ratio, the resultant harmonic-assisted phase amplifications were encouraged forming lower-frequency and higher-frequency bands. The higher-frequency band, which was formed around the beat note via the lateral modal coupling resonance, was pronounced and the phase-amplification bandwidth was increased accordingly.

The collective gain flow from the lower-frequency to the higher-frequency band was shown to trigger the harmonic-assisted phase amplifications in the entire bandwidth based on the intensity circulation analysis. Fifty-fold phase amplification was achieved in the strong feedback regime.

The strong correlation between power spectra and histograms of harmonics-assisted output intensities, which is indicative of the collective nonlinear dynamics in the two frequency bands, has been identified.

We are now pursuing self-consistent theoretical verifications for the detailed experimental results discussed in this work, including the formation of skew-chG mode oscillations with wide-aperture pumping and the resultant nonlinear dynamics responsible for high-degree of harmonic-assisted phase amplifications.